# Fast and deterministic switching of vortex core induced by out-of-plane current in notch disks

Y. M. Luo[1], Y. Z. Wu[2,3], Z. H. Qian[1], J. H. Wen[1], H. Li[1], C. Q. Yu[1], L. Y. Zhu[1] and T. J. Zhou[1]*

1. Center for Integrated Spintronic Devices, Hangzhou Dianzi University, Hangzhou, Zhejiang, 310018, People's Republic of China.
2. Department of Physics, State Key Laboratory of Surface Physics, Fudan University, Shanghai 200433, People's Republic of China.
3. Collaborative Innovation Center of Advanced Microstructures, Nanjing, 210093, China.
*Corresponding to T. J. Zhou (tjzhou@hdu.edu.cn)

**Abstract**

Magnetic vortex, as one of the most interesting magnetic solitons, has attracted great interests in the past two decades. A fast and reliable method to switch vortex polarity and chirality is one of the key issues for various applications. Based on micromagnetic simulation, here we report a fast, low energy cost and deterministic switching of vortex core, through the designing of a notch structure in disks and the use of out-of-plane current geometry. We demonstrate that with such design, the multiple switching problems found in notch disk system can be avoided. Furthermore, the switching time can be reduced by more than 50% compared with disks without notch.

**Introduction**

Magnetic vortex is a flux closure domain structure that exists in a nanodisk of sub-micron size, and can be characterized by an in-plane curling magnetization (chirality) and a nanometer-sized vortex core (VC) with an out-of-plane magnetization (polarity) [1,2]. Due to the clockwise (CW) or counterclockwise (CCW) chirality and the upward ($p=1$) or downward ($p=-1$) polarity, magnetic vortex contains four degenerate states. Magnetic vortex is very stable due to the topological protection. Because of its unique spin structure and dynamic properties, vortices are proposed for various applications. It has been proposed as memory bit for a long time, which have the advantage of store two bits information and high thermal stability[3,4,5,6]. Recent study demonstrated that vortex-based magnetic sensors have the advantage of large linear range, low noise and high sensitivity [7]. Furthermore, vortex-gyration-based spin torque nano-oscillators (STNOs) possess much narrower line width (two-order narrower compared with those utilizing single-domain-procession) [8,9], which can be used as building blocks for neuromorphic computing, being one of promising approach towards low-power artificial intelligence application [10,11]. A full understanding on the reliable control of vortex dynamics and the switching process is one of the outstanding questions for various applications.

In the past decades, great efforts have been made for searching effective methods to control the vortex polarity and chirality. Due to topological protection, a static field about 0.2~0.3 kOe is required to switch the polarity [12]. Further studies find that the polarity can be efficiently switched by using dynamic properties. When excited, the VC would gyrate around the disk center [13], and polarity reversal happens when it reaches a critical velocity. The polarity reversal process is mediated by the formation of vortex-antivortex (VAV) pair [14,15,16]. This efficient polarity switching has attracted great interests [17,18,19]. In addition to magnetic field, VAV mediated polarity switch are also proposed and realized by many other excitations, such as current [20,21], microwave [22] or high-frequency spin wave [23].

However, the limitation of VAV mediated switch process is that to trigger the switch process, VC velocity has to be driven to a critical value, which hinders further optimization on switching speed and energy consumption. One of the methods to overcome such limitiation is to use the boundary effect, where the topological protection can be avoided by moving VC to the edge. It has been reported that in a Pac-man shape disk with an open slot at the disk edge, the polarity can be switched through a dynamic process *via* the nucleation of reversed VC from the notch [24,25,26,27,28]. The energy cost can be greatly reduced, as it does not need to excite the VC to the critical velocity. However, previous results show that such switching has a critical issue: it is very sensitive to the exciting field or current pulse [26, 27], that results in poor reliability for real application. When excite by pulse field, VC switches irregularly with respect to the field amplitude and duration, and an assistant field is needed to help the switching [26]. Current driven VC switching has also been studied for low power and high density device application. Previous studies show that when driven by an in-plane current (CIP configurations), multiple switching (non-deterministic switching) happens. Even though applied current is cut off, vortex core may gyrate back to the notch again, resulting in a second switch process [27]. Therefore, in the notch disk system, to control the polarity, accurate field or current pulse is needed, which is one of the critical issues for real applications. Therefore, searching approaches for better operability and reliability is highly desired.

Previous studies mainly focused on the vortex switching process with in-plane current or field. The VC switching using "current perpendicular to the plane" (CPP geometry) has not been studied yet in notched disks. Compared with the CIP geometry, CPP geometry is more suitable for high density applications. It is known in disks without notch the vortex core can be selectively switched [29,30], through the use of a perpendicular-to-plane polarizer (PERP) in the CPP geometry, allowing to control VC polarity by current direction. Therefore, it is plausible to anticipate that the VC switching irregularity found in notch disks might be solved through the use of PERP in a CPP geometry.

In this paper, based on micromagnetic simulation, we studied the VC switching process in a notched disk using a perpendicularly polarized current, generated in a CPP geometry with a PREP. A fast and reliable switch process with low energy consumption is realized. We show that by using such geometry, the vortex core can be selectively excited and switched. Most importantly, the multiple switch problems that exist in CIP geometry are avoided, resulting from the perpendicularly polarized current that not only excites and switches VC, but also drives the switched VC returning back to the disk center, avoiding the second switching process. Furthermore, such switching is also fast and energetically effective, indicated by the mush shorter switching time, in comparison with disks without notch (more than 150% shorter under low current density). The notch-size-dependent switching diagram indicates a large operation window for such VC switching process. The detailed switching process and the physical origin are fully discussed. Our findings provide an energetically efficient, fast and reliable method to switch VC polarity in notch disks.

## 1. Methods

In the simulation, we study the spin dynamics of the Py nanodisks. The diameter of the disk is fixed at $200\ nm$. The notch depth from the disk edge and the open slot angle were defined as $d$ and $\theta$, respectively. In this study, we choose $d = 80\ nm$, $\theta = 25°$, and the disk thickness $L$ is 40 nm, if not specified. With such a geometry, the ground state is vortex. We choose vortex with CCW chirality and upward polarity as the initial state, as shown in Fig. 1 (b). The current is applied along the z direction. We assume the current is positive (negative) when electrons flow along the –z (+z) direction, and the polarization of the current is out of pane, as shown in Fig.1 (a). The polarized current with perpendicular spin polarization can be realized by passing current through a perpendicular-to-plane polarizer (PERP), such as Co/(Pd, Pt) multilayers [31]. The current has two effects: Firstly, it can excite the magnetization dynamics of the Py layer due to spin transfer torque (STT). Secondly, it can induce a circumferential Oersted field (OH) around the given current pass. Fig.1(c) shows the distribution of OH field produced by current $J = 7 \times 10^6\ A/cm^2$. In our simulation, both effects are considered, and we assume the spin polarization is along –z (+z) direction when positive (negative) current is applied.

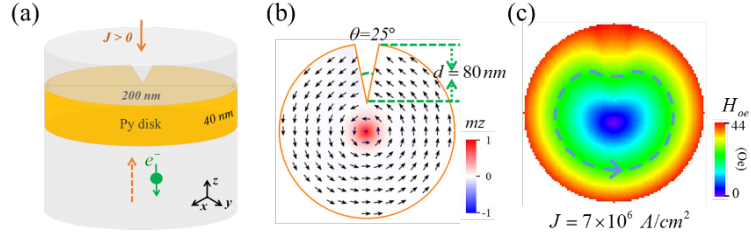

Fig.1 (a) Schematic illustration of the model system. The thickness and diameter of the Py nanodisk are denoted in the figure. The yellow solid and dotted arrow lines denote the direction of current and election flow when positive current is applied, respectively. The spin polarization direction is downward, which is denoted by the arrow. (b) The initial magnetic configuration. The black arrows denote the in-plane curing direction, and the color denotes the normalized out-of-plane magnetization component, as the color bar shows. The size of the notch is denoted in the figure. (c) The spatial distribution of the strength of OH, induced by current flow of the indicated density $J = 7\times10^6$ $A/cm^2$. The arrow and color-coded bar indicate the in-plane rotation sense, and the strength, respectively.

We use OOMMF code to conduct the simulation [32], which employs the Landau–Liftshitz–Gilbert equation [33], with an additional Slonczewski spin transfer torque STT term [34]: $T_{STT} = (a_{STT}/|\overline{M}|)\overline{M}\times(\overline{M}\times\hat{m}_p)$, where $a_{STT} = (1/2\pi)h\gamma PJ/(\mu_0 eM_s L)$, $\hat{m}_p$ is the unit vector of spin polarization direction. $h$ is the Planck's constant, $J$ is the current density, $\mu_0$ is the vacuum permeability, $e$ is the electron charge, $M_s$ is the saturation magnetization, and $P$ is the degree of spin polarization, and here we assume $P = 0.7$ [20,35]. The magnetic parameters of Py is: $\gamma = 2.21\times10^5$ $m/A\cdot s$, $\alpha = 0.01$, $M_s = 8\times10^5$ $A/m$, and the exchange stiffness $A = 1.3\times10^{-11}$ $J/m$. The cell size is $2\times2\times5 nm^3$.

## 2. Results and discussion

Fig.2 shows the VC switching process when a Direct current (DC) $7\times10^6$ $A/cm^2$ is applied. The VC starts to gyrate with increasing amplitude as the current is on. When the VC arrives at the edge of the notch, it moves into the notch and annihilates there, as shown in Fig. 2(a-c). After that, a new VC with reversed polarity starts to nucleate, and finally unpins from the bottom of the notch, as shown in Fig. 2(d-f). To show the details of magnetization dynamic process, in Fig. 2 (f) we plot the time evolution of normalized out of plane magnetization component (Mz) along the notch (the dotted black line in Fig. 2(a)). At 11.47 ns, a positive Mz component is induced at the bottom of the notch as the VC move into the notch. Then the positive Mz disappears as the VC annihilates there. Following that, a new core with opposite polarity nucleates from the bottom of the notch, which is indicated by the formation of negative Mz component at the notch. Finally, the negative Mz component disappear as the new VC unpins from the notch.

The red line in Fig.2 (h) shows the evolution of averaged in-plane magnetization $<m_x>$ as a function of time. The $<m_x>$ oscillates due to the gyration motion of the VC. Therefore the oscillation of $<m_x>$ is able to directly reflect the VC dynamics. The oscillation period of $<m_x>$ is same as VC gyration period, and its amplitude is proportional to VC gyration radius. The increase of oscillation amplitude of $<m_x>$ before 11.7 ns indicates that the VC is gyrating to the notch before switching, while the decrease of oscillation amplitude after 11.7 ns shows that the VC gyrates back to the disk center after switching. The $<m_x>$ time evolution shows that it takes about 7 ns for the switched VC to gyrate back to the disk center after switching. If we switch off the current as soon as the new VC unpins from the notch (at 11.7 ns), the VC would freely gyrate back to the disk center, and it takes about 30 ns for the VC return to the disk center, as shown by the black dash line in Fig. 2(h). This comparison clearly shows that the current not only excites the dynamic motion of VC, but also drives the switched VC returning back to the disk center.

The excitation of VC before switching and the damping after switching is due to the change of the STT direction. The spin polarized current exerts a force on the VC. According to the rigid vortex model, it can be expressed as: $\overrightarrow{F_{ST}} = 2\pi a_{STT} L M_s p / \gamma \overrightarrow{e_\chi}$, where $\overrightarrow{e_\chi}$ is the unit vector tangential to the orbit of the VC [30]. The direction of $\overrightarrow{F_{ST}}$ is determined by the relative direction of the spin polarization $P$ and VC polarity $p$. The $\overrightarrow{F_{ST}}$ is along the damping direction when $P$ and $p$ are parallel, while it is in the opposite direction when $P$ and $p$ is antiparallel. With the fixed current direction ($J > 0$ and $P = 1$), $\overrightarrow{F_{ST}}$ changes from anti-damping to damping direction after VC is switched ( $p = 1 \to -1$ ). That is the reason why the STT is able to excite and switch the VC, and further push the switched VC returning to the disk center after switching.

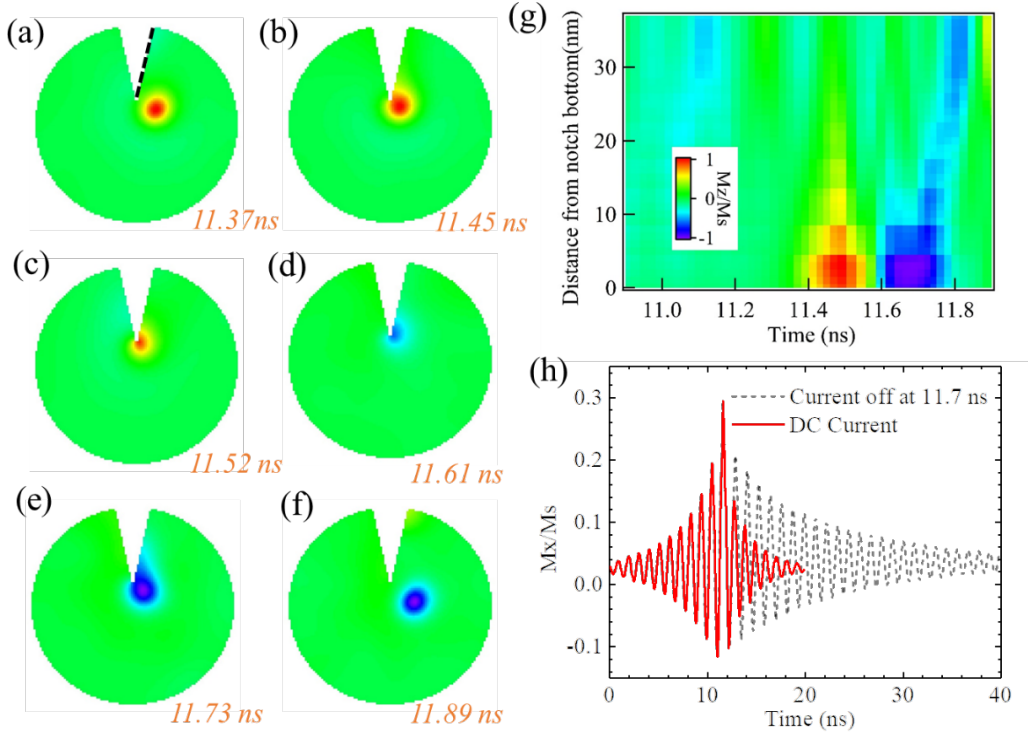

Fig.2 VC switching process at the excitation current of $J = 7 \times 10^6 \, A/cm^2$. (a-f) Selected snapshots of the Mz distribution in the disk during the switch process. (g) The evolution Mz line profile along the edge of the notch (black dot line in (a)). The color in (a-g) represent the normalized Mz direction, indicated by the color bar in (g). (h) The time evolution of $<m_x>$, with current on and off after the VC switching.

    Previously, when an in-plane current is applied, due to the lack of the damping torque after switching, the reversed VC may gyrate back to the notch again, resulting in multiple switching (non-deterministic switching), which has been one of the main problems for practical applications. The simulation results clearly show that the damping force provided by the out of plane spin polarized current can be used to overcome the multiple switching problems that exist in notch disk system. In addition, the quick relaxation to ground state after switching shortens the time interval in favor

of high-frequency successive writing processes.

We then discuss the physical origin of the switching process. For the Py disk geometry chosen in the present study, vortex state is the only stable configuration. Other magnetic configurations such as the C configuration and single domain state are unstable. This ensures the nucleation of a new VC after the annihilation of the original one. To elucidate the VC switching process, we introduce a gyrofield, which is accompanied with magnetiztion gyration dynmics. Previous results haves shown that the gyrofield is critically important in the VC dynamic reversal process, not only for VAV-pair mediated reversal [14,17], but also for notch-induced VC reversal process [27]. The amplitude of the gyrofield is proportional to the VC velocity. Its out of plane component (hz) can be expressed as $hz = -1/(\gamma M_s^2)[\vec{M} \times d\vec{M}/dt]_z$ [36,37], which is opposite to the VC polarity and thus is able to induce VC switching. We calculated hz during the switching process. Figs. 3 (a-f) show the selected snapshots of hz distributions during the switching process, which are corresponding to the magnetization configurations in Figs. 2 (a-f), respectively. *hz* is not spatially uniform and appears in pair with opposite directions around the physical VC. However its negative component is larger than the positive component (Fig. 1(b)), which is consistent with that in circular disks without notch [17]. The amplitude of *hz* decreases at first as VC approaches and gets into the notch. However it increases and reaches the maximum value during the nucleation of the new VC. The decrease of *hz* at first is due to the energy barrier at the notch edges, which retards and slows down the motion of VC when it approaches and gets into the notch, as shown in Figs. 3 (a-c). The kinetic energy of the gyrating VC is high enough to overcome the barrier, and the VC will eventually move into the notch, resulting in the VC annihilation Fig. 2(a)-(f)). The magnetization dynamics clearly show the excitation of spin waves during the original VC annihilation process, while spin waves become weaker and finally disappear during the nucleation of the new VC. The annihilation of orginal VC and the excitation of spin waves indicate that VC gyration energy is converted into spin wave energy, while the disappearing of spin waves and the nucleation of new VC imply that the spin wave energy is converted back to VC gyration energy, leading to the increase of hz during new VC nucleation process, as shown in Fig.3(d) and (e). The negative component of gyrofield around the notch can reach about 1 kOe in amplitude, which is strong enough to determine the reversed VC polarity [24]. As the new VC unpins from the notch, the damping STT torque slows down its gyration speed, leading to the decrease of hz, as shown in Fig.3 (e). The new VC finally reach the disk center, where the gyrofield is around zero.

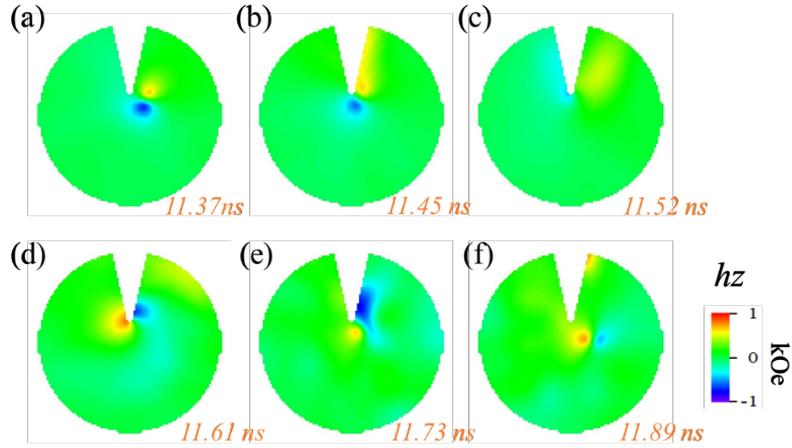

Fig. 3 Selected snapshots of the gyrofield (*hz*) distributions during the polarity switching process，which are correspond to the magnetic configurations in fig.2. (a-f). The excitation current is fixed at $J = 7 \times 10^6 \ A/cm^2$.

**Compare with disk without notch**

To show the advantages of such switching process, we simulate the VC switching process in disk without notch structure for comparison. In circular disk without notch, to induce VAV mediated switching, the critical velocity of VC switching is 360 m/s, and the gyration radius is about 55 nm, while in notched disk, VC switching happens when it reaches the notch and has a speed of 250m/s. As magnetic vortices with upward and downward polarities have opposite skyrmion numbers, the VC switching process can also be denoted by the change of skyrmion numbers [38,28]. The inset of Fig. 4 shows the evolution of skyrmion numbers as a function of time in disk with and without notch, under the same current density of $J = 7\times10^6\ A/cm^2$. The skyrmion numbers change from -1/2 to +1/2 due to the polarity switching. The switch time is 11.7 ns for notched disk, while it is 25 ns for circular disk. The switching time is reduced by 53.2%. Fig. 4 further shows the switching time as a function of current density for disks with and without notch, which clearly shows that the switching time in notched disk can be greatly reduced, and the reduction is more obviously at low current region.

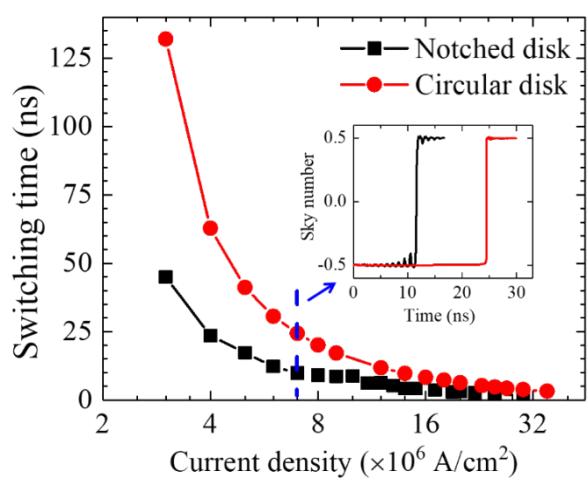

Fig.4 Comparison of switching time as a function of current density in notched and circular disks. The insert show the evolution of skyrmion numbers with time in circular (red) and notched (black) disk when excited by current $J = 7\times10^6\ A/cm^2$.

**The effect of thermal effect, notch size on the VC switch process**

We further show the effect of the notch size on the switching process. Figure. 5(a) show the switching phase diagram as a function of notch depth $d$ and the open slot angle $\theta$, with fixed current density $J = 7\times10^6\ A/cm^2$. The stepsize of $d$ and $\theta$ is 5 nm and $5°$, respectively. According to different switching mechanism, the diagram can be devided into four phases, which can be denoted as I, II, III, IV. The notch induced switching happens in phase II, where $d$ is between 60 to 90 nm, and $\theta$ is below 45°. This wide window for notch medated switching can be readily realized by using moden nano fabrication technique. In phase I, where d is below 60 nm, only VAV-pair mediated switching happens, as the notch is too short and the switching happens before VC reachs the notch. In phase IV, where d is larger than 90 nm and the vortex state is not stable. Therefore new VC is not able to nucleate after the annaihiation of the original one, and the disk finally evloves into the C configuration. Beiside the notch depth $d$, the switching process also depends on $\theta$. The notch boundary forms a repulsive potential that increases with the notch open angle $\theta$ and slows down the motion of the VC As a result of that, the VC may not be able to move into the notch if $\theta$ is too enough. Phase III shows that when $\theta > 35°$, VC switching doesn't happen. Fig. 5(b) shows the detailed VC dynamic process with notch size $d = 70\,nm$ and $\theta = 45°$. As the VC approaches to the notch, it doesn't move into the notch. Instead, the VC changes its gyration direction slightly, moves around the notch top, and finally leaves the notch without annihialtion. Fig. 5 (c) shows the evolution of $<m_x>$ and skyrmion number as a function of time. The skyrmion number is kept negative, indicating that no switching happens. The $<m_x>$ oscillation amplitude increases before 12ns, while keeps the same after that, showing a persistent vortex oscillation, that indicates a dynamically stable state without any polarity switching.

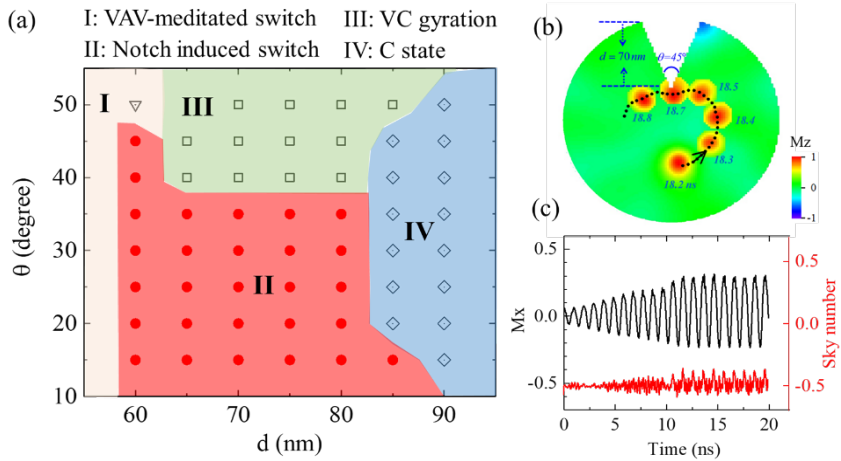

Fig.5 (a) Phase diagram of dependence of switching mechanisms on notch depth ($d$) and angle ($\theta$).The diagram can be divided into four phases as denoted as I, II, III and IV, respectively. (b) and (c): Detailed dynamic process in disk with notch depth $d = 70\,nm$ and angle $\theta = 45°$. (b) The VC positions during the time period of 18ns ~ 19 ns. The red circle represents the VC position at selected time. The black dot line represents the VC trajectory. The black arrow represents the moving direction of the VC. (c) The evolution of the normalized in-plane magnetization component (Mx) and skyrmion number with time.

In addition to the notch size, we also examine the influence of the disk thickness on VC switch process. The simulation results show that the VC can nucleate from the notch and realize switching when the thickness is above 40 nm, below that the disk would transform into the C state. When the disk is thinner than 40 nm, the vortex state is not the only stable configuration. The C state could also exist as a metastable state. In addition, we also consider the thermal effect on the switching process. We simulate the switching process at 300 K, which shows the influence of thermal effect is negligible due to the large size of the vortex. The dependence of notch size, disk thickness, and thermal influence shows the robust and operability of such VC switch process.

## 3. Conclusion

In summary, we studied the VC switching process in a notched disk in a CPP geometry, using out of plane current with perpendicular polarization. Our study provides a low energy cost, good operability and reliable VC polarity switching process. Compared with the switching process using CIP geometry, there are several advantages using such approach: Firstly, The VC polarity can be selectively switched, depending on the current direction. Secondly, such VC switching is deterministic and the multiple switching that exists in CIP geometry is avoided. Lastly, we demonstrate that the switching time can be drastically reduced compared with circular disk without notch. We also study the influence of the notch size and disk thickness on the switching process, showing the operability of such VC switch process. Our work provides a low energy cost and reliable VC polarity switching scheme.

## Acknowledgments

This work was supported by the National Natural Science Foundation (Grants No. 11604066, No. 11874135, No. 61741506, No.11604132, No. 61805061) of China, and Zhejiang Science and Technology Program Project (2017C31061, 2013C31073).